\documentclass[a4paper,11pt]{article}
\usepackage{a4}
\usepackage{amsmath}
\usepackage{amsfonts}
\usepackage{graphicx}
\DeclareGraphicsExtensions{.eps,.eps.gz,.eps.Z,.jpg,.pdf,.bmp}
\usepackage{times}
\usepackage[T1]{fontenc}
\usepackage{indentfirst}

\title
{Adaptive Optics Concept For Multi-Objects 3D Spectroscopy on ELTs}
\author
{B. Neichel$^{1,2}$, T. Fusco$^2$,
  M. Puech$^1$, J-M. Conan$^2$, M. Lelouarn$^4$, E. Gendron$^3$, \\F. Hammer$^1$,
 G. Rousset$^3$, P. Jagourel$^1$ and P. Bouchet$^1$}
\date{}

\begin{document}
\maketitle

{\small
\begin{center}
$^1$GEPI, Observatoire de Paris-Meudon, 5 Place Jules
  Janssen 92195 Meudon, France\\
email : benoit.neichel@obspm.fr \\
$^2$ONERA, BP 72, 92322 Chatillon Cedex, France\\
$^3$LESIA, Observatoire de Paris, 5 Place Jules
  Janssen 92195 Meudon, France \\
$^4$ESO-European Southern Observatory, Karl-Schwarzschild-Stra$\beta$e 2,\\
Garching, D-85748 Germany\\
\end{center}}

\begin{abstract}In this paper, 
we present a first comparison of different Adaptive Optics (AO) concepts to reach a given
scientific specification for 3D spectroscopy on Extremely Large Telescope (ELT).
We consider that a range of $30\%-50\%$ of Ensquarred Energy (EE) in H
band ($1.65$$\mu$m) and in an aperture size from $25$ to $100$mas is representative of the
scientific requirements.
From these preliminary choices, different kinds of AO concepts are
investigated :
Ground Layer Adaptive Optics (GLAO), Multi-Object AO (MOAO) and 
Laser Guide Stars AO (LGS).
Using Fourier based simulations we study the performance of these AO systems depending on the telescope
diameter.
\end{abstract}

% \firstsection % if your document starts with a section,
\section{Introduction}
The new era of astronomical telescopes with diameters reaching 30 to 100m will
provide a dramatic advance in our understanding of the universe.
The concept of ELTs is unique to complement the detection which
will be made from Space. By accommodating high spectral resolution, 3D
spectroscopy devices and relatively large fields of view (10 arc minutes),
ELTs will be a privileged tool for the study of formation and evolution
of galaxies.\\
Adaptive Optics enables large telescopes to provide diffraction
limited images by real time correction of turbulence.
Because the contamination by the interstellar medium light is one of the main
issue in extragalactic studies 
it is required to observe in a direction far from our
galactic plane. In that case, the density of stars becomes dramatically small,
and because of anisoplanatism effects, classical AO working on Natural Guide
Star (NGS) can not be used. To
overcome this problem, new AO techniques (GLAO, MOAO)
have emerged in the last few years to increase the corrected
field using NGS. These new methods are based on a full measurement of the
$3$D turbulent volume using several NGS.
In other hand and to improve sky coverage, Laser Guide Stars (LGS) have been
proposed \cite{Foy85}; \cite{Lelouarn98}. Should some critical issues be solved (cone effect,
spot elongation, tilt indetermination) this last solution would allow the
ultimate scientific requirements to be met.\\
In section 2 we
present the specifications required to achieve scientific goals imposed by
astrophysical studies of high redshifted galaxies.
Section 3 is then dedicated to a first performance evaluation for AO systems,
and some preliminary results from Fourier based simulations will be shown.

% \firstsection
\section{Extragalactic requirements for ELT}\label{sec:ScientificGoal}
% \subsection{Scientific goals}
One of the main science case for ELTs is the physics of galaxies at very high
redshift \cite{Hook04}. Extragalactic studies will benefit from the large
capabilities of ELTs in light concentration
and spatial resolution: 3D spectroscopy of galaxy up to $m_{AB}=26-27$ is expected.
Such data will have a large impact in our understanding of the assembly of dark and
visible matter (from $z=0$ to $z=5$), the physics of galaxies near the
reionisation ($z=6-9$) and the search for the primordial galaxies ($z>>6$).\\
\\
% \subsection{Scientific specifications}
3D spectroscopy of distant galaxies will necessitate to
obtain spatially resolved spectra of very faint and physically small
objects.\\
From a spectroscopic point of view, it will be needed to
resolve velocities of a few
tenths of kilometers, and to avoid contamination between
atmospheric OH lines. This implies to use a spectral resolution
$R$$> 5000-10000$. \\
From a spatial point of view, if we consider a typical galaxy redshifted at
$z$=$6$, its size (half-light radius $R_{half}$) will be around $R_{half}$ =
$0.15$ arcsec \cite{Bouwens04}. Preliminary results \cite{Puech05} from
GIRAFFE (3D spectrograph at the ESO-VLT) show that, in
dynamically unrelaxed systems, the gas is far more extended than the stars:
a pixel size around $R_{half}$ can be sufficient to study galaxy dynamics.
Thus, to be able to resolve different areas (HII
regions, SN, ...) and correctly sample
high redshifted galaxies, the spatial resolution element has to be between
$25$ to $100$ mas. In addition a large FOV will be required to avoid cosmic variance and obtain a
statistically unbiased sample. Scientific goal would require a multiplex factor
ranging from $10$ to $100$
objects, which implies a total FOV of a least $5$ arcmin in diameter. Only new AO
concepts could offer such an angular resolution over such a wide field.\\
Finally, in a first approach, Assemat (2004) \cite{Assemat04} show that a light coupling of
$30-50$\% (in a spatial element of resolution of $250$mas, H band) is required to
detect the $H_{\alpha}$ line with a sufficient SNR (greater than $3$), in a reasonable
exposure time and for $8$m class telescopes. In first approximation we will
preserve this range of light coupling. Once more, AO is essential to increase
the concentration of light within an even smaller spatial element of resolution and meet the
scientific requirements. \\
In the following study, we have fixed the scientific specifications at
$40$\% of light coupling (H band) in a resolution element of $25$ or $50$mas.
This could be subject to future adjustments, but seems
to be a reasonable work baseline.

% \firstsection
\section{Investigating AO solutions}
\subsection{Description of AO systems}

% \begin{itemize}

% \item 
The objective of GLAO is to provide a wide and uniformly corrected field,
but with only very partial correction \cite{Hubin-p-05}; \cite{Rigaut01}. 
This can be perform by compensating for the boundary layer of the
atmosphere which is in the same time the location of most atmospheric
turbulence and the layer per which correction remains valid on a wide FOV.
Several NGS (and several Wave Front
Sensors (WFS)) have to be used to extract
the boundary layer signal from the whole turbulence volume. This can be done
using a simple average of data coming from all the NGS.
Then, only one Deformable Mirror (DM), usually conjugated to the telescope
pupil, is used to correct this ground layer and obtain a uniform correction in the FOV.
We have studied here the performance of a GLAO correction in a total
FOV of $2$arcmin in diameter. That implies that each point of the $2$arcmin FOV see the
same correction (see sect. \ref{results} for results).\\

% \item 
Instead of compensating the whole field, MOAO performs the correction
locally on each scientific object \cite{Hammer01}. Several off axis NGS (a
constellation), even widely outside the isoplanatic angle, are considered to perform a
tomographic measurement of all the turbulence volume around each scientific
object (ie. direction of interest). 
The optimal correction is then deduced
from the turbulence volume estimation and applied
using a single DM per direction of interest.
We focus this study on a MOAO
system working with a NGS constellation of $2$arcmin in diameter around each
scientific object (see sect. \ref{results}).\\

% \item 
To overcome sky coverage issues Laser Guide Stars could be used
where the density of stars becomes too small for good turbulence analysis even
with GLAO or MOAO concepts. But
LGS suffers from limiting effects (Tip-Tilt indetermination, focus
anisoplanatism, spot elongation) which makes
their implementation difficult, especially when Large or Extremely Large
Telescopes are considered. We study here the possibility of using a LGS
with a partial tip-tilt correction. In this case, we suppose a LGS in the direction of
interest providing a perfect correction up to the DM cut-off frequency except
for tip-tilt which is corrected at different levels : correction of $80\%$,
$50\%$ or $0\%$ (no correction) in variance. A partial correction of Tip-Tilt could be achieved if a NGS
is weak or far from optical axis for complete Tip-Tilt
determination.
No other LGS effects (focus
anisoplanatism, spot elongation) were considered. However, previous studies
have shown that focus
anisoplanatism can be solved for instance by using several LGS per object in order
to sense the whole cylinder turbulence path \cite{Viard02}. Concerning spot
elongation issue, customized CCD could be used \cite{Beletic04}.
% \end{itemize}

\subsection{Simulation parameters}
For GLAO and LGS mode, we have developed a simulation tool based on Fourier
approach \cite{Jolissaint06} which computes an AO corrected PSF for different AO systems
\cite{Conan05}. MOAO simulations were done with CIBOLA,
an analytical AO modeling tool developed by B. Ellerbroek \cite{Ellerbroek04}.\\
In GLAO simulation, we assume to measure the average phase over a full
$2$arcmin diameter disk. The pupil layer is then fully corrected up to DM
cut-off frequency. The upper layer correction cut-off is reduced as the
altitude increases (and derived from the WFS FOV). 
LGS simulations are based on classical AO simulations (correction on-axis)
with a partial tip-tilt correction. 
Finally, MOAO simulations assume a tomographic measurement of turbulence thanks to
three stars at the edge of the FOV ($2$arcmin in diameter) and a perfect correction up to DM cut-off frequency.
For the three systems, the DM cut-off frequency is determined by the inter
actuator distance fixed at $0.5$m.\\
Concerning the turbulence, we suppose a typical Paranal turbulence profile, discretized in 10
layers. The total seeing is $0.85$arcsec at $0.5\mu$m and outer scale of
turbulence is fixed at L0=$50$m.\\
Note that some limitations such as WFS noise, temporal error or aliasing are
not taken into account. Results presented here have therefore to be considered as limit (and
for sure optimistic) cases.

\subsection{Simulations results}\label{results}

\begin{figure}[h!]
\begin{center}
     \begin{tabular}{ll}
  \includegraphics[width = 0.5 \linewidth]{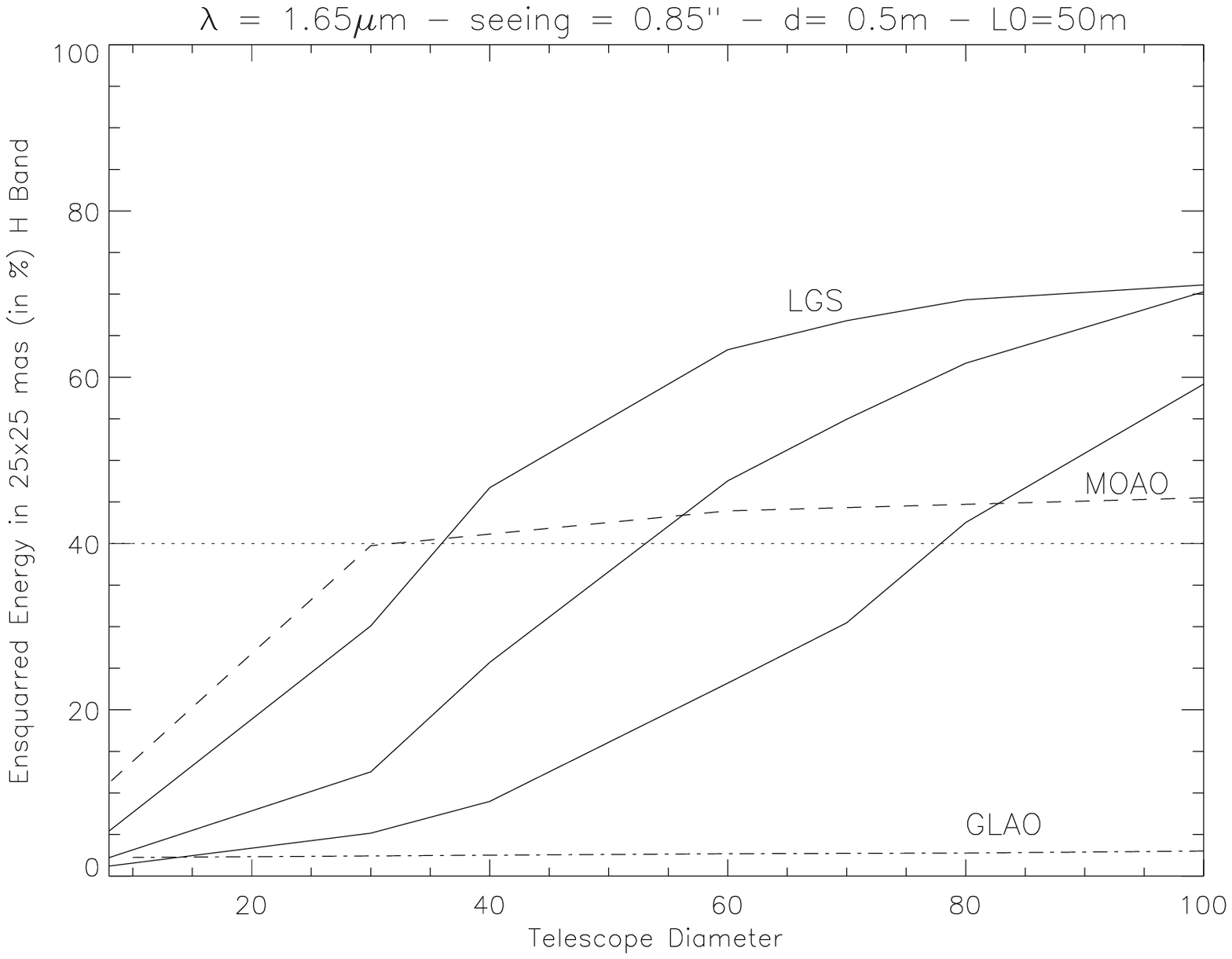} &
 \includegraphics[width = 0.5 \linewidth]{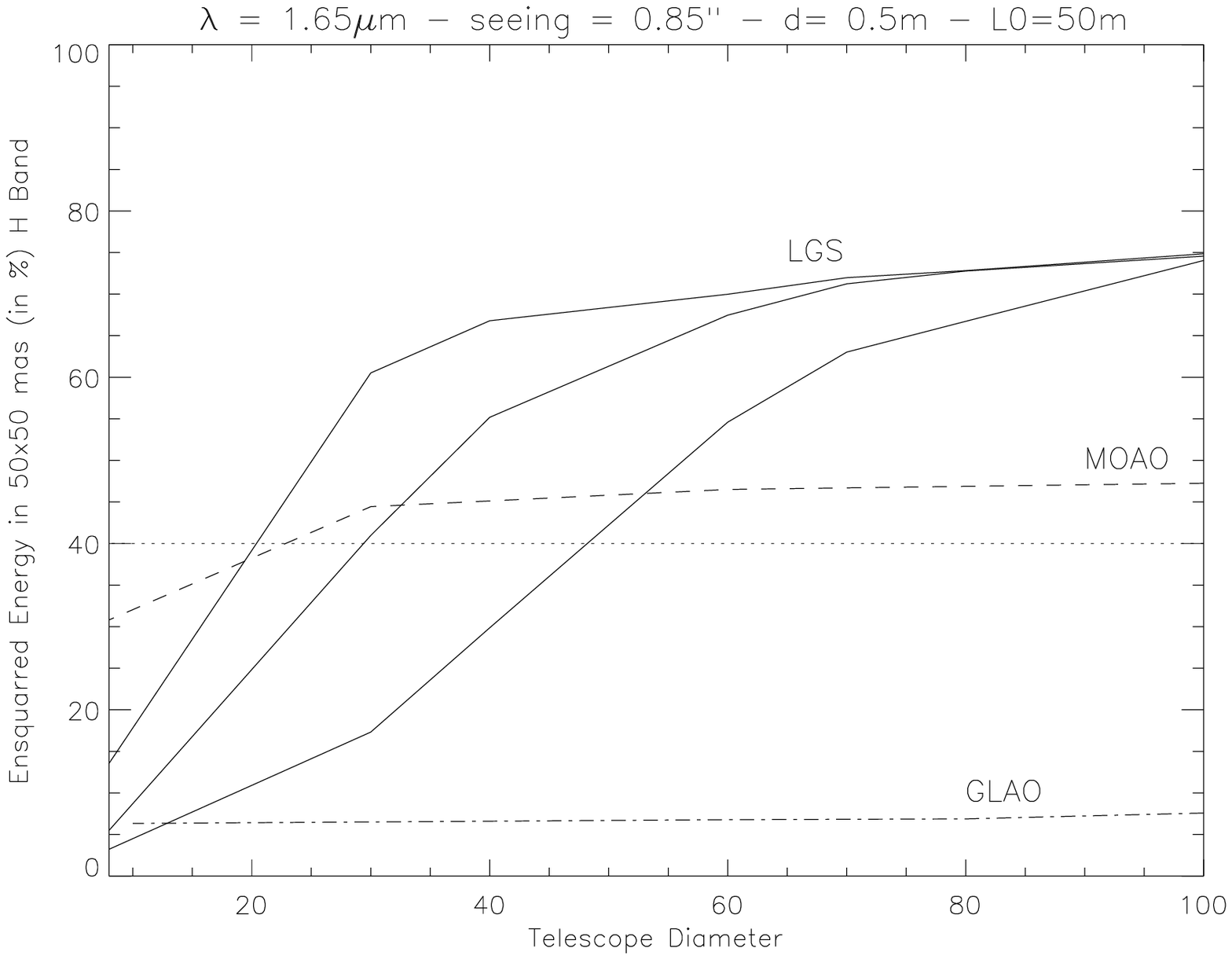} 
 \end{tabular}
\caption{Effect of the telescope diameter (from 8 to 100m) and resolution element size
  (right = 25mas, left = 50mas) on performance for three AO
  systems : GLAO (dashed-dot line) with total FOV=2arcmin, MOAO (dashed line)
  for constellation diameter = 2arcmin and LGS (solid line,
  from top to bottom : 80\% - 50\% - 0\% Tip-tilt correction).}
\label{comp}
\end{center}
\end{figure}

The estimated performance for the three AO systems are presented in figure
\ref{comp}. We plot the Ensquarred Energy (EE) in a given resolution element ($50$ or $25$
mas) for different telescope diameters. We recall that the scientific
specification considered here is to reach $40\%$ of EE, which corresponds to a
gain of a few orders of magnitude compared to the uncorrected case. Seeing
limited case leads to EE$\sim 0.1\%$ in a resolution element of $25$ mas and
EE$\sim 0.7\%$ in $50$mas.\\

Simulations show that :
\begin{itemize}
\item Even in little field ($2$arcmin) the GLAO based system could not reach the
  scientific specifications and performance stay very limited (EE$<10\%$).
  GLAO provides a typical gain of $10$ in
EE compared to seeing limited case, which is not
consistent with the required performance. 
\item Scientific specifications could be reached in a MOAO concept particularly
  when diameter is large (D$>30$m). Indeed, as we consider a fixed resolution
  element, performance benefits from diffraction effects : the size of the PSF
  decreases as diameter becomes larger, and so, the concentrated light in a
  given resolution element increases. For telescope diameter smaller than $30$m, the PSF
  becomes large compared to the resolution element size with the effective
  correction obtained in a MOAO configuration.
\item Using LGS, scientific specifications are achieved when the diameter is large
  enough to overcome the tip-tilt effect or if a NGS is close enough to ensure
  a good tip-tilt correction. Like in the MOAO case, LGS benefits from
  diffraction effects as telescope
  diameter becomes larger. For small diameters, the impact of the resolution
  element size is all the more critical that we consider a partial tip-tilt
  correction. More precisely, it seems that an important parameter could be the ratio
 $D/L0$ multiplied by the resolution element size.
 Indeed, we see that when this parameter becomes
greater than $50$mas, scientific requirements are achieved even without
tip-tilt correction. In other cases, using a NGS even weak
or far from optical axis for tip-tilt measurement should be sufficient to reduce the
residual tip-tilt variance at acceptable levels.
\end{itemize}

\subsection{Discussion on sky coverage}
To perform a correct analysis of turbulence, MOAO needs at least three stars
around the direction of interest. Depending on the position and magnitude of
these stars, performance will change. In other hand, the probability to find a
suitable constellation decrease as we observe at high galactic latitudes
($b>45^\circ$). It has been shown in a previous study \cite{Assemat04}
that a constellation of $2$arcmin in diameter (as the one used to perform the
simulations) could provide $50\%$ sky
coverage at $b=30^\circ$ with $R=16$ stars (respectively $b=60^\circ$ with
$R=17$). But as we consider a perfect case (no noise) for these simulations, we except much
lower performance in a more realistic configuration. Sky coverage could be a
critical issue for MOAO systems working with NGS.\\
In the LGS case we assume a partial tip-tilt correction. That implies to use a NGS for
tip-tilt determination. But as we consider only a partial tip-tilt correction,
the probability to find an adequat NGS is large even in fields far from
galactic plane.\\
If scientific specifications require an aperture size lower than $50$mas and EE$>40\%$, LGS will
become mandatory for $100$\% sky coverage.

% \firstsection
\section{Conclusion}
We have presented here some recent results concerning AO concepts for futurs
ELTs and particularly for 3D spectroscopy.
First numerical simulations show that even for quasi-ideal case, GLAO could
not achieve the scientific goal, MOAO could be used with a limited sky
coverage and
only a LGS based system could offer a 100\% sky coverage.
This article also shows that the implementation of rather simple LGS
systems without Tip-tilt correction could be of interest specifically for very
large telescope diameters.\\ 
Additional simulations are mandatory to precise the science specifications and evaluate
if coupling or aperture size could be relaxed. Further analysis should then
lead to a detailled AO system definition.

\end{document}